\documentclass[11pt]{article}
\usepackage[T1]{fontenc}
\usepackage[utf8]{inputenc}
\usepackage[margin=1in]{geometry}
\usepackage{amsmath}
\usepackage{amssymb}
\usepackage{graphicx}
\usepackage{booktabs}
\usepackage{array}
\usepackage{authblk}
\usepackage[numbers,sort&compress]{natbib}
\usepackage{hyperref}
\usepackage{xurl}

\title{Why pyrotechnics markets keep killing:\\
a simple geometric argument for redesign}

\author[a]{Carlos M. Hern\'andez-Su\'arez\thanks{Corresponding author: \texttt{cmh1@cornell.edu}}}
\author[a]{Alonso S\'anchez-Maldonado}
\author[a]{Carlos A. Robles-Hern\'andez}
\affil[a]{Coordinaci\'on General de Investigaci\'on Cient\'ifica, Universidad de Colima, Bernal D\'iaz del Castillo 340, 28045 Colima, Colima, Mexico}

\date{}

\begin{document}
\maketitle

\begin{abstract}
Fires and explosions in pyrotechnics retail markets recur worldwide with predictable regularity, killing dozens to hundreds of people in single events. This paper argues that the global topology of the market is the dominant determinant of mortality, acting through two independent geometric channels. The first, propagation, concerns ballistic dispersal of ignited articles: the probability that fire spreads between blocks scales with the spatial density of blocks within the dispersal range. The second, evacuation, concerns the distance an occupant must traverse to reach the perimeter, which is set by the global geometry of the market footprint, not by any stall-level parameter. Because mortality risk grows approximately exponentially in evacuation time, topology amplifies modest differences in egress distance into large differences in casualties. Current standards in the United States, the European Union, and Mexico prescribe local parameters such as aisle width and stall separation, but leave the global topology of the market unregulated. We argue that topology should be a regulable design variable, and propose a market geometry that simultaneously slows propagation and shortens evacuation, derived from contact-process models of seed dispersal in spatial ecology.
\end{abstract}


\bigskip

\section{Introduction}

This paper concerns fire and explosion accidents at points of sale
where pyrotechnics are bought by the public, as distinct from
manufacturing or transport accidents. Five catastrophic events
define the reference class: the 1988 explosion at the
La~Merced market in Mexico City, Mexico (at least $62$
dead)~\cite{cenapred1988merced}; the Mesa Redonda fire in Lima,
Peru (2001, $\sim\!280$ dead)~\cite{arce2008mesa}; the
Mercado~Hidalgo explosion in Veracruz, Mexico (2002, $30$
dead)~\cite{cenapred2003veracruz}; the Puttingal temple disaster in
Paravur, India (2016, $110$ dead)~\cite{parkash2016puttingal,
illiyas2018routine}; and the recurrent fires at the San~Pablito
market in Tultepec, Mexico (most lethal event in 2016 with $42$
dead, four destructive episodes between 2005 and
2018)~\cite{cenapred2016sanpablito}.

Existing standards regulate inventory and inter-magazine
distances~\cite{atf555}, local stall parameters such as aisle
width and supervision~\cite{nfpa1124}, the conformity of the
articles themselves~\cite{eudirective2013}, and store-level
containment and suppression~\cite{leon2023}. A parallel literature
treats accident causation in pyrotechnic
manufacturing~\cite{nallathambi2023} and threshold quantities for
storage legislation~\cite{wharton2004}; egress in general retail
is well-developed within the Required Safe Egress Time / Available
Safe Egress Time (RSET/ASET) framework~\cite{kuligowski2008}. None
of this regulates the \emph{topology} of the market as a whole.

We argue that this topology is in fact the dominant geometric
determinant of mortality, acting on two independent channels.
\emph{First, propagation.} Under ballistic dispersal of ignited
projectiles---the regime that applies to pyrotechnics---the
probability that a fire propagates from one block to another is
governed by the spatial density of blocks within the market
footprint. A compact two-dimensional layout surrounds every
burning block with other blocks in every direction, so that an
ignited rocket has a high probability of landing on flammable
material; a linear arrangement places blocks along a single thin
strip, so that most directions of dispersal lead to vacant
exterior. This is a direct corollary of the \emph{basic contact
process} with seed dispersal~\cite{hernandez2012}: the probability
of sustained spread is a monotone increasing function of the
density of receptive sites within the dispersal range. \emph{Second,
evacuation.} Once a fire has started, survival depends on the
distance an occupant must traverse to reach the perimeter.

The remainder of the paper is organized as follows.
Section~\ref{sec:model} derives the evacuation distance under two
idealized topologies. Section~\ref{sec:regulation} surveys the
major regulatory regimes governing pyrotechnics retail and shows
that all three converge on the same topology-blind structure.
Section~\ref{sec:discussion} synthesises the propagation and
evacuation channels, situates the gap within the framework of
nested socio-technical control, and outlines an implementation
pathway. Section~\ref{sec:conclusion} concludes.

\section{Two idealized layouts and the evacuation distance}
\label{sec:model}

\begin{figure}[t]
\centering
\includegraphics[width=\linewidth]{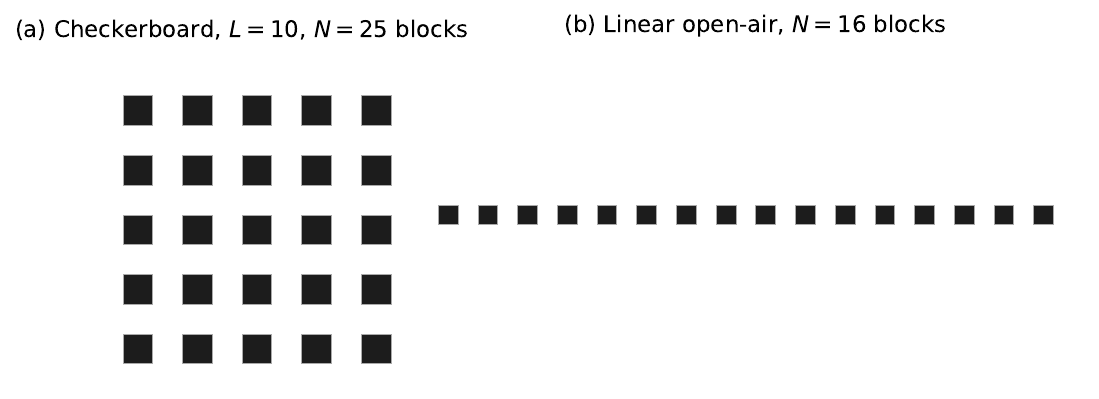}
\caption{The two idealized market layouts. (a)~A compact
\emph{checkerboard} on an $L \times L$ grid, representative of the
prevailing footprint of pyrotechnics retail markets worldwide.
(b)~A \emph{linear open-air} layout, with long sides directly open
to the exterior. In the linear arrangement every aisle cell is one
step away from safety, regardless of $N$. In both panels $N$
denotes the number of \emph{blocks} (panel (a) shows $N=25$ blocks
at $L=10$; panel (b) shows $N=16$ blocks).}
\label{fig:layouts}
\end{figure}

We compare two idealized market layouts holding the number of
blocks $N$ constant (Fig.~\ref{fig:layouts}). The
\emph{checkerboard} layout places blocks on a regular sub-lattice
of an $L\times L$ grid with spacing two: every block is surrounded
by eight aisle cells (a Moore neighborhood of
aisles~\cite{hernandez2012}). With $L$ even, $(L/2)^2$ blocks fit
on the grid, so $N=L^2/4$ and $L=2\sqrt{N}$. The \emph{linear
open-air} layout is a single row of $2N$ cells alternating block
and aisle, with both long sides open to the exterior. The
``blocks'' in our model are arranged like city blocks in an urban
grid: each block is a compact cluster of physically adjacent
retail stalls separated from neighboring blocks by aisles, but
with negligible separation between stalls within the same block.
Once any stall in a block ignites, the entire block burns as a
unit. For San~Pablito (Fig.~\ref{fig:sanpablito}), the market
comprised approximately $300$ individual retail stalls organized
as $80$ blocks of about four stalls each, on a grid of
$8\times 10$ blocks. Each occupant evacuates along the shortest
Manhattan path to the perimeter; the performance metric is the
expected Manhattan distance from a uniformly chosen aisle cell to
the nearest exterior point, a deliberately conservative proxy for
the Required Safe Egress Time (RSET) that ignores crowding, smoke,
and panic and assumes occupants move towards the closest edge.

\begin{figure}[t]
\centering
\includegraphics[width=\linewidth]{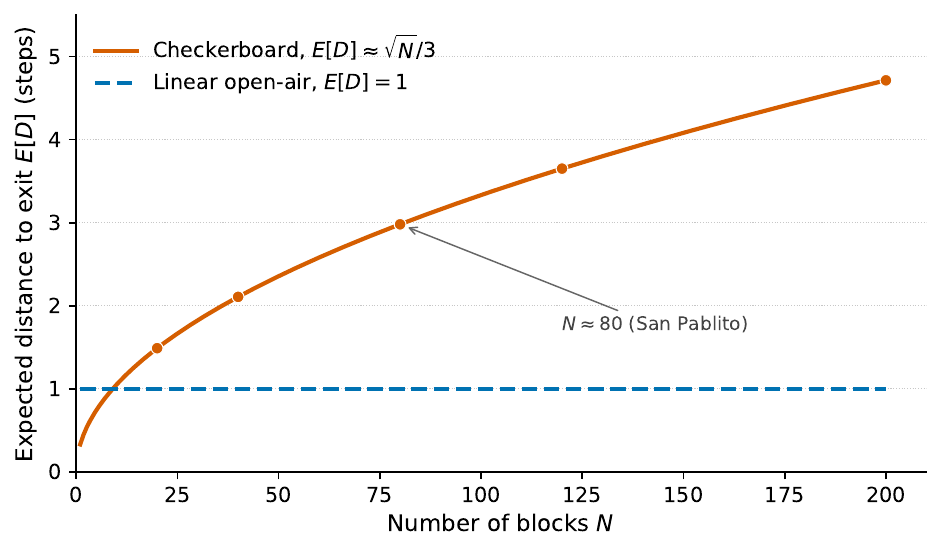}
\caption{Expected Manhattan distance to the perimeter (in
block-widths) as a function of the number of blocks $N$. The
checkerboard penalty grows as $\sqrt{N}/3$ while the linear layout
stays at one block-width.
Mortality risk grows approximately exponentially in evacuation
time, so the ratio in expected casualties is substantially larger
than the ratio in distance plotted here.}
\label{fig:scaling}
\end{figure}

Throughout, distance is measured in \emph{block-widths}: one
block-width equals the lattice spacing of two grid cells (one block
plus one adjacent aisle), so that a single step from an aisle cell
to the exterior counts as one unit. Take a place at random inside
the checkerboard design and let $D$ be the distance (in
block-widths) to the closest edge. It is possible to show that
$\Pr(D\ge k) = ((L-2k)/L)^2$ for $k=0,\ldots,L/2-1$, and by summing
the tail probability and approximating the sum by a Riemann
integral, one obtains $E[D]\approx L/6$; substituting
$L=2\sqrt{N}$,
\begin{equation}
E[D_{\text{checkerboard}}] \approx \frac{\sqrt{N}}{3}.
\label{eq:checkerboard}
\end{equation}
The formula is asymptotic; the exact tail sum gives
$E[D] = L/6 - 1/2 + 1/(3L)$, so for moderate $N$ a constant
correction of $-1/2$ applies. In the linear layout every aisle cell
sits one step from the exterior, so $E[D_{\text{linear}}] = 1$ in
the same unit. The escape
distance in the checkerboard grows without bound as $\sqrt{N}$,
while in the linear layout it is constant; for a market of
$N\approx 80$ blocks (San~Pablito), the linear layout is
approximately three times shorter to escape
(Fig.~\ref{fig:scaling}).

\section{Comparative regulatory analysis}
\label{sec:regulation}

The major regulatory regimes governing pyrotechnics retail---the
United States, the European Union, and Mexico---differ in scope,
structure, and enforcement, but converge in one striking respect:
all three regulate \emph{local} parameters in detail while leaving
the \emph{global} topology of the retail market entirely
unspecified.

In the United States, NFPA~1124~\cite{nfpa1124} governs the
construction, location, and operation of retail sales facilities
for consumer fireworks, prescribing limits on stored quantities,
inter-stall separations, fire-resistant construction, and on-site
supervision. ATF~27\,CFR~555~\cite{atf555} sets minimum
quantity-distance tables for the storage of display fireworks.
Both documents specify the geometry of \emph{individual} units
(stalls, magazines, storage rooms) with precision but are silent
on how those units should be collectively arranged into a market.

In the European Union, Directive~2013/29/EU~\cite{eudirective2013}
(the recast of Directive~2007/23/EC) harmonises the placing on the
market of pyrotechnic articles across Member States, mandating
conformity assessment by notified bodies, four fireworks categories
(F1--F4) with corresponding minimum age limits, and the essential
safety requirements of Annex~I. The directive regulates the
\emph{product} comprehensively---design, labelling, traceability,
distribution---but leaves the layout of the retail venue to
national law, which in practice means to local building codes that
do not address pyrotechnic risk specifically.

In Mexico, a cluster of Normas Oficiales Mexicanas (NOM)
regulates the manufacture, transport, and storage of pyrotechnics,
primarily through the Secretar\'ia de la Defensa Nacional
(Sedena) for permitting and through state-level civil-protection
authorities for storage and retail. Despite the recurrence of
mass-casualty events---most prominently the 2016 San~Pablito
explosion (Fig.~\ref{fig:sanpablito})---no provision constrains
the global geometric arrangement of retail stalls beyond
inter-magazine distance requirements that are routinely contested
or waived on grounds of cultural heritage and economic
displacement.

This pattern is not an accident of jurisdiction. A Safety Science
contribution that surveyed eleven national regulatory regimes for
the Dutch Ministry of Social Affairs~\cite{wharton2004} found the
same focus across countries: threshold \emph{quantities} (typically
expressed in TNT-equivalent kilograms) and \emph{inter-installation
distances} between magazines, but no parameter describing the
topology of a retail venue once those individual constraints are
met. The reform proposed there, like the reforms that have
followed in other jurisdictions, raised or lowered numerical
thresholds while leaving the structural variable untouched.

Table~\ref{tab:regs} summarises the comparison. Across all three
major jurisdictions, the topological level of the market footprint
falls outside the scope of any regulation that mentions
pyrotechnics specifically; it is governed only by generic urban
zoning that does not consider the propagation and evacuation
dynamics distinctive to this hazard.

\begin{table}[t]
\centering
\small
\caption{Scope of pyrotechnic-specific regulation in three major
jurisdictions. ``\checkmark'' indicates the dimension is
regulated; ``--'' indicates it is unaddressed at the level of
pyrotechnic-specific code.}
\label{tab:regs}
\begin{tabular}{@{}p{0.36\linewidth}ccc@{}}
\toprule
 & US & EU & MX \\
\midrule
Product certification    & \checkmark & \checkmark & \checkmark \\
Storage quantities       & \checkmark & \checkmark & \checkmark \\
Inter-magazine distance  & \checkmark & \checkmark & \checkmark \\
Stall construction       & \checkmark & --         & \checkmark \\
Aisle width              & \checkmark & --         & --         \\
\textbf{Market topology} & \textbf{--} & \textbf{--} & \textbf{--} \\
\bottomrule
\end{tabular}
\end{table}

\section{Discussion}
\label{sec:discussion}

\begin{figure}[t]
\centering
\includegraphics[width=0.72\textwidth]{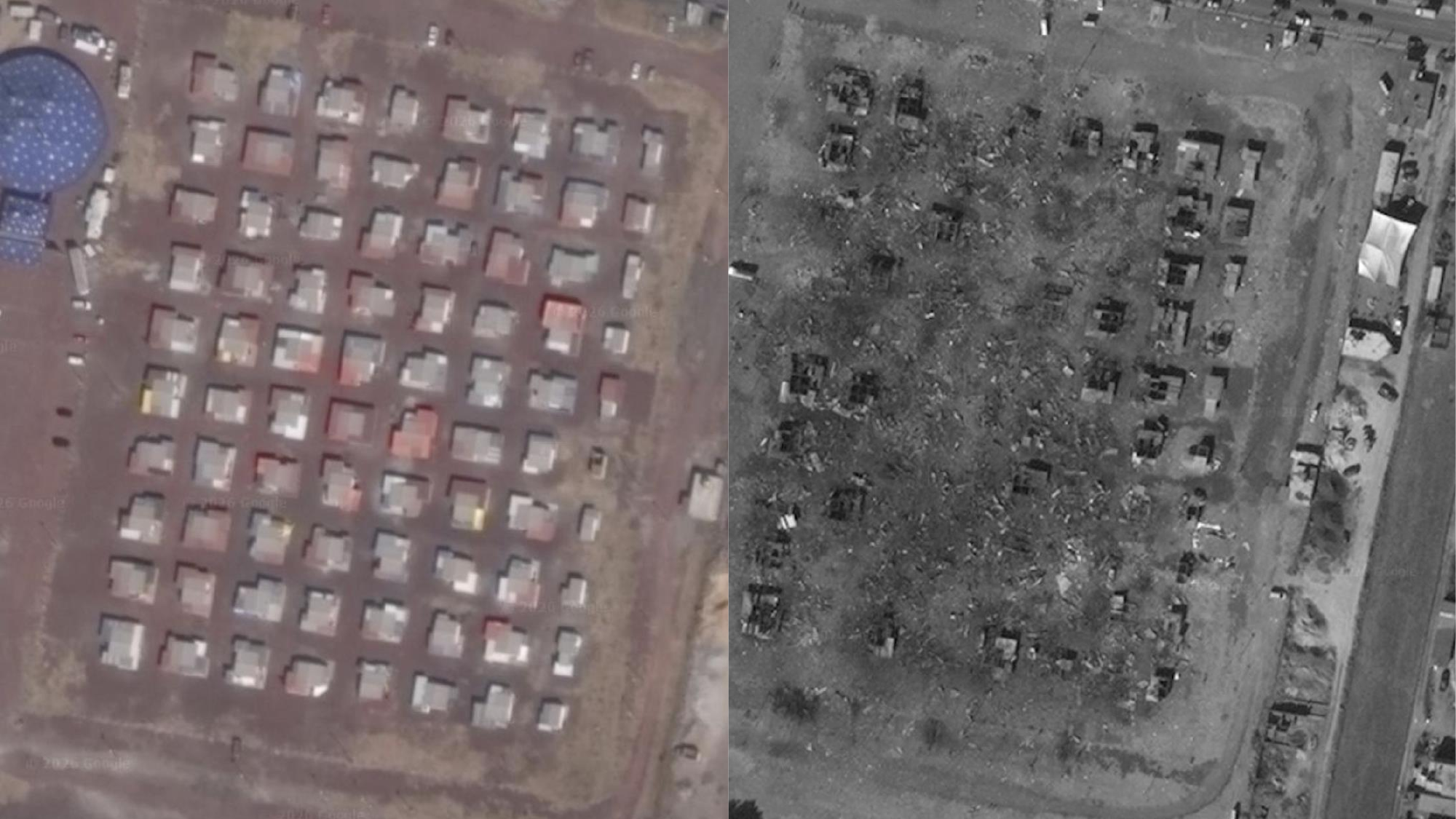}
\caption{Satellite imagery of San~Pablito Market, Tultepec, Mexico
($19^\circ40'06''$~N, $99^\circ07'37''$~W), before and after the
fireworks explosion of 20~December~2016 that killed 42~people.
Left panel: the market in normal operation; the regular
two-dimensional arrangement of approximately 80~blocks of stalls
is clearly visible. Right panel: the same area two days after
the explosion; nearly the entire grid of blocks has been
destroyed. Imagery: Google Earth (web), captured 23~January~2016
and 22~December~2016.}
\label{fig:sanpablito}
\end{figure}

Let $\gamma$ be the probability that a burning block sustains
active combustion for one more time step, so that low $\gamma$
corresponds to fast burnout. Within the BCP
framework~\cite[Table~1]{hernandez2012}, the critical $\gamma$
required for sustained propagation is approximately $25\%$ higher
in a linear arrangement than in a Moore configuration. The
mechanism is geometric: counting receptive blocks within a
Chebyshev dispersal radius $r$ (in block-widths) of a burning
block, a block in a checkerboard is surrounded on all sides while a
block in a line has receptive neighbors only along the spine. At
$r=3$ a block in a checkerboard has $8$ times more receptive
neighbors than a block in a line; at $r=5$, $12$ times more. The same
combustion conditions that sustain a fire in a checkerboard can
extinguish it along a line.

The factor-of-three advantage in expected distance is in fact a
lower bound on the advantage in mortality. Under fire conditions,
the probability of incapacitation grows approximately exponentially
in evacuation time, because toxic-gas dose, heat-flux exposure,
and the probability of being overtaken by propagating fire all
accumulate multiplicatively. The same point applies to behavior:
the derivation assumes that each occupant walks the shortest path
to the perimeter, an assumption that favors the checkerboard
rather than the linear layout. Under realistic
conditions---smoke obscuring sightlines, herding, panic, blocked
aisles---paths are longer; but in the linear arrangement every
direction (other than back along the strip) reaches safety almost
immediately, whereas in the checkerboard a disoriented occupant
can wander deeper into the interior. Both mortality non-linearity
and behavioral departure from optimality widen the gap between
the two layouts.

The San~Pablito case (Fig.~\ref{fig:sanpablito}) embodies the
prediction. The pre-incident image shows the canonical compact
two-dimensional topology---approximately eighty blocks of four
stalls each, arranged on an $8\times10$ grid, with aisles too
narrow and too few to reach the perimeter in a single
direction---and the post-incident image shows the predicted
outcome: a near-total destruction of the grid by propagation
across receptive neighbors, with the worst casualties concentrated
in the interior cells where the evacuation distance was longest.

The pattern identified here---that pyrotechnic regulation
constrains parameters at the level of the individual stall and
the individual storage container, but is silent at the level of
the market as a whole---is precisely the kind of gap that the
nested-control framework of socio-technical risk
management~\cite{rasmussen1997} predicts and warns against. Risk
in a complex system is managed through nested levels of control:
legislators set policy, regulatory agencies translate policy into
standards, market authorities translate standards into local
rules, and individual vendors implement those rules at the stall
level. When one level of this nested structure is not represented
in any control instrument, the system as a whole has no mechanism
to constrain choices at that level. In the case of pyrotechnics
retail, the architectural-topological level---the layout of the
market footprint---falls into precisely this gap. Each individual
stall complies with applicable rules; each storage container
respects required distances; each operator carries the required
permit. Yet the market as an emergent assembly of compliant units
acquires properties (compact two-dimensional topology, exponential
evacuation distance) that no single rule addresses, and which no
inspector is empowered to challenge.

The relevant property of the linear layout is not its strict
one-dimensionality but the fact that the width of the block
arrangement perpendicular to its principal direction is $O(1)$,
independent of $N$. This admits a family of variants that preserve
constant $E[D]$ while accommodating real-world constraints: two
parallel rows separated by a wide central aisle, a comb
arrangement with shallow perpendicular branches off a linear
spine, a hollow rectangle whose interior serves as a safety zone,
or a curved or zigzag strip folded to fit urban frontage. The
regulatory rule we propose should be read as identifying a class
of admissible topologies, not prescribing a single architectural
form.

The minimum implementable change is a clause requiring that new
pyrotechnics retail markets above a threshold size (for example,
$N\ge 50$ blocks) be configured so that the expected Manhattan
distance from any aisle position to the exterior does not exceed
a stated bound such as two block-widths. This is a topology-level
constraint that admits the family of variants above while ruling
out compact two-dimensional designs at any non-trivial size. It is
straightforward to verify at the permitting stage from a site
plan; it requires no instrumentation, training, or behavioural
change once enforced; and its implementation cost is essentially
the cost of refusing permits for non-compliant layouts.

Several limitations of the analysis deserve emphasis. First, the
derivations assume idealised topologies; real markets are
partially constrained by streets, terrain, and historical
buildings, and a clause requiring linearity would be ineffective
if it could not be enforced through retrofitting or relocation.
Second, the propagation argument rests on ballistic dispersal as
the dominant transfer mechanism; convection-driven spread in
confined corridors could in principle reverse the topological
preference, although in the historical cases the proximate cause
has consistently been projectile-mediated ignition rather than
thermal radiation alone. Third, the mortality estimate depends on
the assumption that walking speed and exposure tolerance are
uncorrelated with topology, which is a simplification: panic-
induced behavior may differ between narrow linear corridors and
open grids. None of these caveats reverses the central comparison;
each suggests directions for empirical refinement.

A fourth caveat concerns the two channels jointly. The evacuation
and propagation benefits of linearity are not independent once the
strip is folded to fit an urban footprint. The folded variants that
preserve constant $E[D]$---the comb, the zigzag, the hollow
rectangle---bring segments of the spine back into spatial
proximity, so that blocks far apart along the strip may fall within
ballistic dispersal range of one another. Folding therefore retains
the evacuation advantage, which depends only on the $O(1)$
perpendicular width, while partially eroding the propagation
advantage, which depends on keeping receptive blocks out of
dispersal range. The unfolded strip maximises both benefits
simultaneously; practical folded layouts trade some propagation
resistance for compactness. A complete treatment would optimise the
fold geometry against the dispersal radius $r$, ensuring that the
spacing between adjacent spine segments exceeds $r$; we leave this
to a design-specific analysis.

A fifth caveat concerns which channel dominates the death toll. We
have treated propagation and evacuation as separate additive
effects that both favour linearity, but their relative contribution
to mortality is not established empirically. In events such as
San~Pablito a substantial fraction of deaths plausibly arises from
the initial sympathetic detonation and immediate fire spread rather
than from failed egress, in which case the propagation channel
dominates and the exponential-in-egress-time argument---on which the
``factor of three is a lower bound on the mortality advantage''
claim rests---carries proportionately less weight. Relatedly, the
critical-$\gamma$ result quantifies the threshold for
\emph{sustained}, percolating spread, whereas a mass-casualty
outcome does not require percolation: a single large detonation
within a compact cluster can be lethal without a self-sustaining
front. The contact-process threshold is therefore suggestive of the
direction of the effect rather than a direct predictor of casualty
counts. Disentangling the two channels would require incident-level
reconstruction of where and when fatalities occurred relative to
the propagating front, which the available aggregate records do not
support.

\section{Conclusion}
\label{sec:conclusion}

The global topology of pyrotechnics retail markets is the dominant
geometric determinant of mass-casualty outcomes, acting both on
the probability that fire propagates and on the distance victims
must traverse to escape. Both effects favor linear configurations
over compact two-dimensional ones. The architectural-topological
level of the market is the one level of the socio-technical
control structure that current pyrotechnic-specific regulation
leaves entirely unconstrained; closing this gap is the minimum
change consistent with the historical record. The argument is
elementary and the implementation cost is modest: roadside artisan
markets, food corridors, and coastal restaurant rows operate on
essentially linear footprints worldwide, demonstrating that the
topology is commercially viable, culturally accepted, and
economically robust at scale. The recurring death tolls justify
acting on it.

\section*{Declaration of generative AI and AI-assisted
technologies in the manuscript preparation process}

During the preparation of this work the authors used Anthropic's
Claude to assist with manuscript drafting, literature search,
bibliography verification, and \LaTeX{} formatting. After using
this tool, the authors reviewed and edited the content as needed
and take full responsibility for the content of the published
article.

\bibliographystyle{unsrtnat}
\bibliography{references_safscience}

\end{document}